\documentstyle[bibnorm]{lamuphys2}
\newcommand{\Zint}{Z}
\newcommand{\Real}{R}
\newcommand{\irrep}[1]{{\bf #1}}
\newcommand{\A}{{\cal A}}
\newcommand{\K}{{\cal K}}
\newcommand{\V}{{\cal V}}
\newcommand{\C}{{\cal C}}
\newcommand{\M}{{\cal M}}
\newcommand{\T}{{\cal T}}

\def\sp{\;\;\;\; , \;\;\; \; }
\def\co{\;\;\; :  \;\;\;\; }

\begin{document}
\title{
U-duality and M-theory, an algebraic approach$^*$
}

\author{Niels A. Obers \inst{1} \and Boris Pioline \inst{2}}

\institute{Nordita and Niels Bohr Institute, Blegdamsvej 17,
DK-2100 Copenhagen, Danmark 
\and
Centre de Physique Th{\'e}orique, {\'E}cole polytechnique, F-91128 Palaiseau, France
}
\maketitle
\footnotetext[1]{To
appear in the proceedings of {\it Quantum Aspects of Gauge
Theories, Supersymmetry and Unification}, Corfu, September 1998. \\ 
\hfill NORDITA-98/78 HE, NBI-HE-98-42,  
CPHT-PC698.1298, \tt hep-th/9812139}

\begin{abstract}
Based on our work \cite{Obers:1998fb}, we discuss how U-duality
arises as an exact symmetry of M-theory from T-duality and 11D
diffeomorphism invariance. A set of Weyl generators 
are shown to realize the Weyl group of $SO(d,d,\Zint)$
and $E_{d(d)}(\Zint)$, while Borel generators extend these finite groups
into the full T- and U-duality groups. We discuss how the BPS states
fall into various representations, and obtain duality invariant mass
formulae, relevant for the computation of exact string amplitudes.
The realization of U-duality symmetry in Matrix gauge theory is also
considered. 
\end{abstract}

\section{Introduction}
\enlargethispage{1.5cm}
Once thought of as distinct ten-dimensional 
consistent perturbative theories, the five superstring theories
are now understood as five different asymptotic expansions
of a unifying eleven-dimensional M-theory in different corners
$g_s \rightarrow 0$
of its moduli space, with {\it string dualities} providing the
transition functions between these
patches\cite{Townsend:1995kk}.
Due to the asymptotic nature of string perturbation theory, this
still only covers a zero-measure set of the moduli space,
while analytic continuation to finite $g_s$ can be achieved
for quantities that are protected from quantum corrections, such as
the BPS spectrum and BPS-saturated amplitudes. One-loop exact
amplitudes on one patch translate into non-perturbative amplitudes
in the dual one, and world-sheet instantons turn into 
semi-classical configurations of Euclidean branes wrapped on
non-trivial cycles of the target space. 

On the other hand, non-perturbative {\it symmetries} identify
different points in the moduli space and hold for any amplitude.
The prototypical example is the $Sl(2,\Zint)$ S-duality symmetry 
of ten-dimensional type IIB theory, which was used to obtain
the exact four-graviton $R^4$ amplitude as the 
weight $s=3/2$ $Sl(2,\Zint)$ Eisenstein series \cite{Green:1997tv}:
\begin{equation}
E_{s=3/2}^{Sl(2,\Zint)} = 2 \zeta(3) \sum_{p \wedge q=1} \alpha^3_{(p,q)}
\end{equation}
where $\alpha_{(p,q)} = \sqrt{\tau_2} \alpha'/| p + q \tau| $ 
is the Einstein-frame inverse tension of a $(p,q)$ string,
and $\tau=a+i/g_s$ the complex modulus in the upper half plane.
This result follows from supersymmetry constraints and S-duality 
invariance \cite{Berkovits:1997pj}. 
The weak coupling expansion reproduces
the tree-level and one-loop contributions, implies perturbative
non-renormalization beyond one-loop, and exhibits a 
sum of D-instanton effects. More generally, 
exact amplitudes should be given by (non-holomorphic)
automorphic forms of the symmetry group, selected on the basis
of the leading perturbative behaviour and supersymmetric
constraints \cite{Kiritsis:1997em}.
From a study of this and other exact results,
one may hope to infer the rules of instanton calculus in string
theory\cite{Green:1997tv,Kiritsis:1997em,Bachas:1997mc}.

The purpose of this talk is to extend these considerations to the
lower-dimensional case of M-theory compactified on a torus $T^d$,
where the symmetry gets enlarged to the U-duality group
$E_{d(d)}(\Zint)$ \cite{Hull:1995mz}, 
and correspondingly more states come into play.
We first review some basic facts about M-theory
and its relation to type IIA string theory, then discuss the
continuous hidden symmetry \cite{Cremmer:1980gs} of toroidally 
compactified 11D supergravity. Only a discrete subgroup 
of this symmetry may remain as a quantum symmetry of M-theory,
and we show how this U-duality follows from T-duality
and 11D diffeomorphism invariance. We first focus on a
set of Weyl generators of these groups, and derive
how the BPS spectrum falls into representations thereof \cite{Elitzur:1997zn}.
Borel generators are then included to obtain duality invariant mass formulae.
Finally, we use the connection between M-theory in 
Discrete Light-Cone Quantization and Matrix gauge theories 
\cite{Banks:1997vh}
to uncover the spectrum of the latter theories,
and discuss the realization of U-duality in these 
non-gravitational theories. 
This talk is a shortened version of Ref. \cite{Obers:1998fb},
to which the reader is referred for a more extensive discussion 
and list of references. 

\section{M-theory, BPS spectrum and hidden symmetries}

M-theory was originally introduced \cite{Townsend:1995kk} as the 
strong coupling limit of
type IIA string theory. It reduces to 11D supergravity at low energies,
and is conjectured to exhibit eleven-dimensional N=1 super Poincar{\'e} 
invariance. In particular, M-theory compactified on a circle $S^1$ of
radius $R_s$ gives type IIA string theory with 
\begin{equation}
\label{matching}
R_s/l_p = g_s^{2/3}\ ,\quad l_p^3 = g_s l_s^3
\end{equation}
where $l_p$ denotes the 11D Planck length, and
$g_s$ and $l_s$ the string coupling and length. Indeed, one observes that
at strong coupling
$g_s \rightarrow \infty$ the eleventh direction decompactifies.
The massless spectrum of type IIA string theory gets unified in the
11D setting as
\begin{equation}
 g_{MN} \simeq ( g_{\mu \nu}, \phi, \A_\mu ) 
\;\;\;\;, \;\;\;\;  
 \C_{MNR} \simeq ( B_{\mu \nu},  \C_{\mu \nu \rho}) 
\end{equation}
in terms of the M-theory graviton and antisymmetric three-form
and the NSNS fields $(g,B,\phi)$ and RR gauge potentials
$(\A_1,\C_3)$ of type IIA string theory (the lower
index denotes the number of antisymmetric indices); 
so do the half-BPS states, 
charged under these fields:
\begin{equation} 
KK =(DO,KK)\ ,\quad M2 = (F1,D2)
\ ,\quad M5 = (D4,NS5) 
\ ,\quad KK6 = (KK5,D6,\dots)
\end{equation}
Here the first (resp. second) entry in each parentheses corresponds to
longitudinal (resp. transverse) reduction with respect to 
the eleventh direction. This identification can be 
most easily seen from the mass or tension of the M-theory BPS states, 
$\T_0 = 1/R_s $,
$\T_2 = 1/l_p^3$, 
$\T_5 = 1/l_p^6$, 
$ \T_6 = R_{TN}^2/l_p^9$
for the Kaluza-Klein momentum state, M2 and M5 brane, and
KK6-brane (or Taub-NUT monopole), using the relations in (\ref{matching}). 
As a side remark, the KK6-brane may also be reduced along a
non-compact direction of the Taub-NUT space, giving rise to 
a state with tension proportional to $R_{TN}^2 R_s/l_p^9 \sim
1/g_s^3$ and a logarithmically divergent geometry at infinity
(first reference in \cite{Blau:1997du}). As we shall see, U-duality predicts 
many more such states with exotic tensions
$\T \sim 1/g_s^{p \geq 3}$ for compactification to $D \leq 3$ dimensions
\cite{Obers:1998fb,Blau:1997du}.

Upon compactification on a torus $T^{d}$, 11D SUGRA exhibits a group
of continuous global symmetry $G$, as well as a (composite) gauge
invariance under its R-symmetry maximal compact subgroup $K$,
as listed in Table 1~\cite{Cremmer:1980gs}. 
The symmetry can be seen among the
scalar fields, which take value in the symmetric space $K\backslash G$.
The local $K$-invariance can be gauge-fixed thanks to the Iwasawa
decomposition
\begin{equation} 
\label{iwa}
 G  = K \cdot A \cdot N
\end{equation} 
into maximal compact $K$, Abelian $A$ and
nilpotent $N$ subgroups. A natural gauge is obtained by taking $K=1$, so that
the moduli  can be represented as a vielbein: $\V \in A \cdot N$
or equivalently as a $K$-invariant matrix $M = \V^t \V$. 
In this representation,  the Abelian factor $A$ is parametrized by the
radii $R_I$ of the internal
torus, while the nilpotent factor $N$ incorporates the expectation values
of the off-diagonal components of the metric and
of the gauge potentials $\C_3$ on the torus, as well as their
Poincar{\'e}-dual $\K_{1;8}$ and ${\mathcal{E}}_6$. The symmetry group $G_d$ acts
on the right on the coset $K\backslash G$, and induces a compensating
moduli-dependent $K$-rotation to preserve the gauge $K=1$.
The massless spectrum also includes a number of $p$-forms,
given for $p=1,2$ in the last two columns of Table 1, where dualization
into forms of lower degree has been carried out. They 
transform linearly under $G$ and induce charges for
particles ($m$) and strings ($n$) respectively. Those are linearly
related to the central charges in the $N=8$ supersymmetry algebra by
$Z=\V \cdot m$, so that the invariant quadratic form 
$$
\M^2 = Z^t Z = m^t M m \sp M = \V^t \V 
$$
is precisely the invariant mass for half-BPS states (quarter-BPS
states receive further corrections).

\begin{table}[h]  
\begin{center}
\begin{tabular}{|r|r||l|l|l|l|l|}
\hline
$D$ & $d$ & $G=E_{d(d)}$ & $K$ & scalars & 1-form & 2-form \\
\hline
10& 1 &$\Real^+$                             & 1& 1 &  \irrep{1} & \irrep{1} \\
9 & 2 &$Sl(2,\Real)\times \Real^+$          &$U(1)$& 4-1 & \irrep{3} & \irrep{2} \\
8 & 3 &$Sl(3,\Real)\times Sl(2,\Real)$&$SO(3)\times U(1)$& 11-4  & \irrep{(3,2)} & \irrep{(3,1)} \\
7 & 4 &$Sl(5,\Real)$                  &$SO(5)$& 24-10 & \irrep{10} & \irrep{5} \\
6 & 5 &$SO(5,5,\Real)$                &$SO(5)\times SO(5)$ & 45-20 & \irrep{16} & \irrep{10} \\ 
5 & 6 &$E_{6(6)}$                     &$USp(8)$& 78-36 & \irrep{27} & \irrep{27} \\
4 & 7 &$E_{7(7)}$                     &$SU(8)$& 133-63 & \irrep{56} & .   \\ 
3 & 8 &$E_{8(8)}$                     &$SO(16)$& 248-120  & .  &  .   \\
\hline
\end{tabular}
\end{center}
\centerline{Table 1: Cremmer-Julia symmetry groups and scalars, 1-forms and 2-forms.}
\end{table}

While the low-energy effective action is invariant under the
continuous group $G$, the existence of charged states implies
that the symmetry remaining at the quantum level is at most
a discrete subgroup thereof. In particular, some charges correspond
to momenta in compact directions (resp. string or brane wrapping
numbers around cycles of the torus), and therefore take values in
the reciprocal (resp. homology) lattice. The most convenient 
constraint takes place in $D=4$, where the quantum symmetry
should preserve the symplectic Dirac quantization constraint
\cite{Hull:1995mz}:
\begin{equation} 
E_{d(d)}(\Zint) \subset E_{7(7)}(\Real) \cap Sp(56,\Zint)\ .
\end{equation}
On the other hand, 11D diffeomorphism invariance implies that
the mapping class group  $Sl(d,\Zint)$ of the torus $T^d$ be
an exact symmetry, whereas T-duality holds to all orders in type IIA string
perturbation theory and therefore in M-theory as well. These
two symmetries certainly preserve Dirac quantization, and the
U-duality group of M-theory is therefore
\begin{equation}
\label{udgr}
E_{d(d)} (\Zint) = Sl(d,\Zint)\bowtie SO(d-1,d-1,\Zint) \ ,
\end{equation}
where the symbol $\bowtie$ denotes the group generated by the two
non-commuting subgroups. In the following, we elucidate this
equality by studying the action of a set of Weyl and Borel generators.
 
\section{Dynkinese approach to T- and U-duality}

We first construct a set of Weyl generators for the T-duality
symmetry group $SO(d,d,\Zint)$ of toroidally compactified type II 
string theory. The moduli space of the 
NSNS sector of type II string theory on $T^d$ displays a
structure analogous to (\ref{iwa}), with now $G=SO(d,d,\Real)$
and $H= SO(d) \times SO(d)$,
where $A$ are the radii of the torus
and $N$ the off-diagonal components of the metric and antisymmetric
tensor. There is also an extra factor $\Real^+$ for the dilaton.
We define the Weyl generators as those preserving a rectangular
torus with zero two-form vev, namely $N=1$. A minimal generating set
is provided by the exchange of radii and a double T-duality on
two directions (a single T-duality would map to type IIB):
\begin{equation}  
\label{tdw1}
S_{i}: R_i \leftrightarrow R_{i+1}\ ,\quad i=1\dots d-1
\end{equation} 
\begin{equation}  
\label{tdw2}
T: (g_s,R_1,R_2) \leftrightarrow \left(\frac{g_s}{R_1 R_2},\frac{1}{R_2},
\frac{1}{R_1}\right)
\end{equation} 

In order to understand the structure of this group, we represent
the tension of a state 
${\mathcal{T}}=g_s^{x^0} R_1^{x^1} R_2^{x^2}\dots R_d^{x^d}$
as a vector $\lambda=x^0 e_0+ \dots x^d e_d$ in a weight space,
on which the generators $S_i$ and $T$ act linearly. In fact the metric
on this space can be chosen so that they are 
orthogonal reflections
\begin{equation} 
\label{weyl}
  \lambda \rightarrow \rho_\alpha (\lambda) =
\lambda - 2 \frac{\alpha\cdot~\lambda}{\alpha\cdot \alpha} \alpha\
,\qquad
(\lambda,\lambda)=-(x^0)^2+(x^i)^2+x^0 x^i
\end{equation} 
with respect to planes normal to the set of simple roots $\alpha_0$, 
$\alpha_i$.
The non-Euclidean metric of the space can be evaded by noting
that the $D=(10-d)$-dimensional Planck length 
$ V_R/g_s^2 l_s^8 $ is invariant, and restricting to the
subspace orthogonal to $\delta=e_1+\dots+e_d-2 e_0$.
The group structure is then determined by the inner products between
roots, summarized in the Dynkin diagram
\begin{equation} 
\label{dynt}
\begin{array}{ccccccc}
& \circ_{0} & \left(\frac{1}{g}\right)  &&&&\\
& \mid          &&&&&\\
\left(\frac{R_1}{g}\right) \;\; \circ_{1} - & \circ_{2}  &-&
\circ_{3} &- \dots-&
\circ_{d-1} \;\; \left(\frac{1}{R_d}\right)  \\
\end{array}
\end{equation} 
which is precisely the one of $SO(d,d)$. On this diagram,
we have also indicated the tensions of the fundamental weights
associated to the the extremal nodes, which show that
the KK modes on the torus together with the string
winding states transform as a vector, while the two spinor representations
correspond to a multiplet of D-particles and D-strings respectively.

We now apply the same techniques to the Weyl group  of the  
U-duality group (\ref{udgr}) of M-theory on $T^d$
\cite{Elitzur:1997zn}. In this case, we
first need to translate the double T-duality (\ref{tdw2}) in M-theory variables
using (\ref{matching}), and conjugating this with an 11D
diffeomorphism $R_s \leftrightarrow R_i$, we find the $\Zint_2$ symmetry
\begin{equation}
T_{IJK} \co 
R_I \rightarrow \frac{l_p^3 }{ R_J R_K}   
\sp 
l_p^3 \rightarrow \frac{l_p^6 }{R_I R_J R_K }  
\end{equation} 
which involves a set of {\it three} directions. A minimal set of
generators of the Weyl group 
can be chosen as $T=T_{123}$ and the permutations 
$S_I : R_I \leftrightarrow R_{I+1}$. 
The action of the U-duality Weyl group is
now represented on a weight vector $\lambda= x^0 e_0 +\dots
x^d e_d$ by considering the tension monomials 
${\mathcal{T}}=
l_p^{3 x^0} R_1^{x^1} R_2^{x^2}\dots R_d^{x^d}$.
The  generators $T$  and $S_I$ can again be implemented as orthogonal
reflections (\ref{weyl}) for the F-theory--like metric
$(\lambda,\lambda)=-(x^0)^2+(x^i)^2$,
with respect to roots $\alpha_0$, $\alpha_I$ with inner products 
\begin{equation}
\label{dynu}
\begin{array}{ccccccccc}
&&&&\circ_{0}&\left(\frac{1}{l_p^3}\right)&&&\\
&&&&\mid& &&&\\
\left(\frac{R_1}{l_p^3}\right) \;\; \circ_{1} &-&
\circ_{2} &-&
\circ_3 &-&
\circ_4&-\dots-&
\circ_{d-1} \;\; \left(\frac{1}{R_d}\right)
\end{array}
\end{equation}
Not surprisingly, this is the Dynkin diagram of the $E_{d(d)}$
groups. Note that for $d=9$, the 
invariant Planck length vector $ V_R / l_p^9$ becomes null,
and it is no longer possible to go to the orthogonal subspace. 
This is consistent with the fact that $E_9$ is the affine Lie group
associated to $E_8$, and implies that the representations are
infinite dimensional. For $d>9$, the symmetry is even more dramatic,
with hyperbolic $E_{10}$ or Kac-Moody $E_{11}$, while for $d<9$
we recover the groups in Table 1. The representations 
corresponding to the extremal nodes indicated 
in (\ref{dynu}) correspond to a {\it particle} multiplet with highest
weight $\M =1/R_d$ (a KK mode) , a {\it string} multiplet with
highest weight $\T_1 = R_1/l_p^3$ (a singly wound M2-brane),
and a {\it membrane} multiplet with highest weight $1/l_p^3$. The
particle and string multiplets are precisely the ones charged
under the one- and two-form potentials in Table 1. The other
members in the same multiplet are obtained by applying Weyl
reflections on these highest weights, and
e.g. for $d=7$,  the particle multiplet transforms in  the \irrep{56}
of $E_{(7(7)}$ with content listed in Table 2.  The multiplet consists
of the particle states obtained from the KK state and completely wrapped
M2, M5 and KK6-branes. It restricts to the lower-dimensional particle
multiplets upon decompactification, and decomposes as perturbative
states, D-branes and NS-branes under T-duality. We do not display the $d=8$ 
particle multiplet, but point out that the latter branches into
additional representations under T-duality, and in particular contains
states with a mass of order $1/g_s^3$, akin to the KK6-brane
discussed in section 2. This behaviour arises as soon as less
than three transverse dimensions are present, and signals a non
asymptotically flat geometry.
\begin{table}
\begin{center}
\begin{tabular}{|c|l|l||c| }
\hline
mass $\M$ & $Sl(7)$ irrep & charge & $E_{\rm YM}$ \\ \hline
$\frac{1}{R_I}$ & \irrep{7}& $m_1$&  $\frac{g^2_{\rm YM} s_I^2 }{N V_s}$   \\
$\frac{R_I R_J}{l_p^3}$ & \irrep{21} &  $m^2 $& $\frac{V_s}{ N g^2_{\rm YM} (s_I s_J)^2 }$    \\
$    \frac{R_I R_J R_K R_L R_M}{l_p^6}$ & \irrep{21} & $m^5$& $\frac{V_s^3}{ N g^6_{\rm YM} (s_I s_J s_K s_L s_M  )^2 }$     \\
$\frac{R_I^2 R_J R_K R_L R_M R_N R_P}{l_p^9}$ 
& \irrep{7} & $m^{1;7}$& $\frac{V_s^5}{ N g^{10}_{\rm YM} (s_I; s_J s_K s_L s_M s_N s_P s_Q  )^2 }$   \\
\hline
\end{tabular}
\end{center}
\centerline{Table 2: Particle multiplet, \irrep{56} of of $E_{7(7)}$ and
 Yang-Mills
 interpretation.}  
\end{table}

\section{Borel generators and invariant mass formulae}

Next, we wish to include the Borel generators, corresponding 
to the right action of nilpotent matrices $N$ in the Iwasawa decomposition
(\ref{iwa}), in order to derive duality invariant mass formulae. 
In the case of T-duality, the Borel generators include Dehn twists
from the mapping class group $Sl(d,\Zint)$, acting on the homology
lattice as $\gamma_I\rightarrow \gamma_I+\gamma_J$, as well as
the spectral flow of the B-field:
\begin{equation} 
\label{tbor}
\Gamma_{ij} \;\; : \;\;\;\; A_j^{(i)} \rightarrow A_j^{(i)} + 1
\sp 
\Delta_{ij} \;\; : \;\;\;\; B_{ij} \rightarrow B_{ij} + 1
\end{equation}
Using the $Sl(d,Z)$ boosts, a KK state with momentum 1 can
be related to a state with any integer momentum $m_i$ along the
circles of $T^d$ with mass $\M^2 = m_i g^{ij} m_j$. More generally,
using (\ref{tbor}), the vector representation 
of the Weyl group of T-duality becomes a
lattice of KK and winding charges $(m_i,m^i)$ with mass
\begin{equation}
{\mathcal{M}}_0^2 = 
 (m_i + B_{ij} m^j)
g^{ik} (m_k + B_{kl} m^l) + m^i g_{ij} m^j 
\end{equation} 
subject to the half-BPS condition
$\|m\|^2 = 2 m_i m^i =0$. For quarter-BPS states the mass formula receives
an extra contribution proportional to $\| m\|^2 $. Similarly, 
the D-particle states (completely wrapped D-branes)
with masses $1/g_s l_s$,$ R_i R_j/g_s l_s^3$, 
$R_i R_j R_k R_l /g_s l_s^5,\ldots$ 
turn into a  lattice of integer 
D0,D2,D4-charges $m = \{ m ,m^{ij}, m^{ijkl}, \ldots \}$
transforming as spinor of $SO(d,d,\Zint)$.  
A careful analysis of the Borel generators then gives the half-BPS mass
\begin{equation} 
 \M_0^2 = \frac{1}{g_s^2} \left[ \tilde m^2 +
 (\tilde m^{ij})^2 +
(\tilde m^{ijkl})^2 +\dots \right]  
\end{equation}
where the $\tilde{m} = {\mathcal{V}} \cdot m $ are the dressed charges 
$\tilde m = m +   m^{ij} B_{ij} +  m^{ijkl}
B_{ij} B_{kl}+\dots,
\tilde m^{ij} = m^{ij} +   m^{klij} B_{kl} + \dots,
\tilde m^{ijkl} = m^{ijkl} + \dots$,
with an extra contribution when 
the half-BPS conditions,
$k^{ijkl}  \equiv  m^{[ij} m^{kl]} + m~m^{ijkl} = 0 $ for $d=4$,
are not fulfilled. These BPS mass formulae
can also be obtained by analyzing the Born-Infeld action 
\cite{Dijkgraaf:1997hk,Bachas:1997mc}. 

In the case of the U-duality group, the Borel generators include the
usual Dehn twists $\Gamma_{IJ}$ in $Sl(d,\Zint)$ as well as the $B_{ij}
=\C_{sij}$ shifts in $SO(d-1,d-1,\Zint)$, which can be conjugated into
$C_{IJK}: {\mathcal{C}}_{IJK} \rightarrow  {\mathcal{C}}_{IJK}+1$.
When $d \geq 6$ we also need to include shifts of the dual
gauge fields ${\mathcal{E}}_6, \K_{1;8}$.  
The U-duality invariant mass formula on tori with arbitrary gauge
backgrounds can then again be obtained by considering the 
non-commuting $\C_3$, ${\mathcal{E}}_6$ and $\K_{1;8}$ flows.
In $d=7$, we obtain the result 
\begin{equation} 
\M_0^2 = \left( \tilde m_1 \right)^2 +
         \frac{1}{ l_p^6} \left( \tilde m^{2} \right)^2 +
         \frac{1}{ l_p^{12}} \left( \tilde m^{5} \right)^2 +
         \frac{1}{ l_p^{18}}  \left( \tilde m^{1;7} \right)^2
\end{equation} 
where the  dressed charges are given by  
\begin{equation} 
\begin{array}{ll}
\tilde m_1  = & m_1 +  \C_3 m^2 + \left( \C_3 \C_3 + {\mathcal{E}}_6 \right) m^{5} 
       +\left( \C_3 \C_3 \C_3 + \C_3 {\mathcal{E}}_6 \right) m^{1;7} \\ 
\tilde m^2  =& m^2 + \C_3 m^5 + \left( \C_3 \C_3 + {\mathcal{E}}_6 \right) m^{1;7} \\ 
\tilde m^5 =& m^5 + \C_3 m^{1;7} \\  
\tilde m^{1;7}= & m^{1;7}  
\end{array}
\end{equation} 
It turns out that the half-BPS condition on the particle multiplet transforms
as the string multiplet constructed out of the particle charges 
\cite{Obers:1998fb}, which also follows on dimensional grounds:
\begin{equation} 
\label{cons}
\begin{array}{lll} 
k^1 &=& m_1 m^2 \equiv 0 \\
k^4 &=& m_1 m^5 + m^2 m^2 \equiv 0 \\
k^{1;6} &=& m_1 m^{1;7} + m^2 m^5 \equiv 0 \\
k^{3;7} &=& m^2 m^{1;7} + m^5 m^5 \equiv 0 \\
k^{6;7} &=& m^5 m^{1;7} \equiv 0 
\end{array}
\end{equation} 
When these composite charges do not vanish, the state is at
most quarter-BPS, in which case its mass formula reads 
$
{\M}^2 = {\mathcal{M}}^2_0(m) + \sqrt{ \left[ \T(k) \right]^2 } 
$
with  $\T(k)  $  
the half-BPS tension formula for the string multiplet and $k$ the
quadratic charges in (\ref{cons}).  

\section{U-duality in Matrix gauge theory}

Finally, we wish to use the information on U-duality and invariant
mass formulae obtained above to obtain a better understanding of
the degrees of freedom and symmetries of matrix gauge theory.
According to this prescription 
\cite{Banks:1997vh}  
the DLCQ of M-theory on a lightlike circle
$S^1$ (radius $R_l$) times $T^d$ in the sector with $P_+ = N/R_l$ is
described by a supersymmetric $U(N)$ gauge theory in $d+1$ dimensions
on the maximal T-dual torus $\tilde{T}^d$.
In this description the  KK state of the particle multiplet bound 
to $N$ D0-branes becomes  (after maximal T-duality) 
a string wound on $N$ D$d$-branes with YM energy  
$E_{\rm YM} =  g_{\rm YM}^2 s_I^2 / N V_s = \M^2 / P_+ $   
describing a half-BPS state carrying electric flux. 
On the other hand, a wrapped membrane of the string multiplet bound to  
$N$ D0-branes
becomes  
a KK state bound to $N$ D$d$-branes with energy  
$E_{\rm YM} = 1/s_I = R_l \T_1 $  
describing a massless excitation. More generally, the 
U-duality invariant gauge theory
masses  can be computed as 
\begin{equation} 
E_{\rm YM} = 
\frac{\M^2 }{P_+} + R_l \T_1 
\end{equation} 
As an example, the last column of Table 2 gives the corresponding
gauge theory masses corresponding to the particle multiplet of
$E_{7(7)}$. The first state is the state carrying electric flux
mentioned above, while the second state carries magnetic flux, and
corresponds in general to a D$d$-D$(d-2)$ bound state.  
In the case of the string multiplet, the details of which we omit here,
the first states are the KK excitations, and the second is a 
YM instanton in 3+1 dimensions lifted to $d +1$ dimensions,
corresponding to D$d$-D$(d-4)$ bound states. Starting on $T^5$,
we see the occurrence of states with energy $1/g_{\rm YM}^4$
whose interpretation is problematic in pure Yang-Mills terms.

We can make a more precise comparison of the U-duality invariant masses with
the proposed quantum descriptions for $d=3,4,5$. In the first
case the D3-brane description is valid so that we have N=4 
supersymmetric gauge theory in 3+1 dimensions. The U-duality
symmetry $Sl(3,\Zint) \times Sl(2,\Zint)$ 
corresponds to the mapping class group of the
three-torus together with the S-duality symmetry of the gauge theory
\cite{Susskind:1996uh}.
In particular, the M-theory or type IIB  modular parameter
$\tau =  \C_{123} + i V/l_p^3$ is equated to
$S = \theta/2 \pi + i 4 \pi/g_{\rm YM}^2$, so that
the M-theory gauge potential $\C_{123}$ translates
as a $\theta$ angle in the gauge theory.
For $d=4$ the quantum theory is the (2,0) worldvolume theory on the
M5-brane, where the extra dimension (D4 $\rightarrow$ 
M5) is generated by identifying 
$R_5 = g_{\rm YM}^2$. 
The $Sl(5,\Zint)$ U-duality symmetry is now naturally interpreted as the
symmetry of M5 on the five-torus, and indeed from the invariant masses
we can infer directly  that $\A^I = \epsilon^{IJKL} \C_{JKL}$
appears as the KK gauge field on $T^5 = T^4 \times S^1$.
Finally, for $d=5$ the quantum theory is the non-critical string theory
living on the NS5-brane (obtained by type IIB S-duality from the D5-brane) 
with string length $\hat{l}_s^2 = g_{\rm YM}^2$.  The $SO(5,5,\Zint)$ U-duality
symmetry is now interpreted as the T-duality symmetry of the string
theory, and from the invariant masses we infer that  
$ B^{IJ}  = \epsilon^{IJKLM} \C_{KLM}$  
is the $B$-field of string theory on NS5-brane. 
For  $d\geq6$ gravity does no longer decouple anymore. 
These conclusions can also be reached from the Born-Infeld action 
in the scaling limit \cite{Obers:1998fb}. 
Finally, we mention that the string and
particle multiplet of $E_{d(d)}(\Zint)$, together with the momentum
$N$ and some additional charges, make a string multiplet 
representation of $E_{d+1(d+1)}(\Zint)$, as should be the case
if the DLCQ proposal is correct \cite{Obers:1998fb,Blau:1997du}.

\enlargethispage{2cm}
\providecommand{\href}[2]{#2}\begingroup\raggedright\endgroup


\begin{thebibliography}{10}

\bibitem{Obers:1998fb}
N.~A. Obers, B.~Pioline, and E.~Rabinovici {\em Nucl. Phys.} {\bf B525} (1998)
  163, \href{http://xxx.lanl.gov/abs/hep-th/9712084}{{\tt
  hep-th/9712084}}~;
N.~A. Obers and B.~Pioline \href{http://xxx.lanl.gov/abs/hep-th/9809039}{{\tt
  hep-th/9809039}}, to appear in Phys. Rept.

\bibitem{Townsend:1995kk}
P.~K. Townsend {\em Phys. Lett.} {\bf B350} (1995) 184,
  \href{http://xxx.lanl.gov/abs/hep-th/9501068}{{\tt hep-th/9501068}};
\\
E.~Witten \href{http://xxx.lanl.gov/abs/hep-th/9507121}{{\tt hep-th/9507121}}.

\bibitem{Green:1997tv}
M.~B. Green and M.~Gutperle {\em Nucl. Phys.} {\bf B498} (1997) 195,
  \href{http://xxx.lanl.gov/abs/hep-th/9701093}{{\tt hep-th/9701093}}.

\bibitem{Berkovits:1997pj}
N.~Berkovits \href{http://xxx.lanl.gov/abs/hep-th/9709116}{{\tt
  hep-th/9709116}};
B.~Pioline {\em Phys. Lett.} {\bf B431} (1998) 73,
  \href{http://xxx.lanl.gov/abs/hep-th/9804023}{{\tt hep-th/9804023}};
M.~B. Green and S.~Sethi \href{http://xxx.lanl.gov/abs/hep-th/9808061}{{\tt
  hep-th/9808061}}.

\bibitem{Kiritsis:1997em}
E.~Kiritsis and B.~Pioline {\em Nucl. Phys.} {\bf B508} (1997) 509,
  \href{http://xxx.lanl.gov/abs/hep-th/9707018}{{\tt hep-th/9707018}};\\
N.~A.~Obers and B.~Pioline, work in progress.

\bibitem{Bachas:1997mc}
C.~Bachas, C.~Fabre, E.~Kiritsis, N.~A. Obers, and P.~Vanhove {\em Nucl. Phys.}
  {\bf B509} (1998) 33, \href{http://xxx.lanl.gov/abs/hep-th/9707126}{{\tt
  hep-th/9707126}};
B.~Pioline and E.~Kiritsis {\em Phys. Lett.} {\bf B418} (1998) 61,
  \href{http://xxx.lanl.gov/abs/hep-th/9710078}{{\tt hep-th/9710078}};
P.~Vanhove, 
  \href{http://xxx.lanl.gov/abs/hep-th/9712079}{{\tt hep-th/9712079}};
B.~Pioline, 
\href{http://xxx.lanl.gov/abs/hep-th/9712155}{{\tt hep-th/9712155}}.

\bibitem{Hull:1995mz}
C.~M. Hull and P.~K. Townsend {\em Nucl. Phys.} {\bf B451} (1995) 525,
  \href{http://xxx.lanl.gov/abs/hep-th/9505073}{{\tt hep-th/9505073}}.

\bibitem{Cremmer:1980gs}
E.~Cremmer, B. Julia in {\em Superspace and
  Supergravity}, S.~W. Hawking and M.~Rocek, eds.
\newblock Cambridge University Press, 1981.

\bibitem{Elitzur:1997zn}
S.~Elitzur, A.~Giveon, D.~Kutasov, and E.~Rabinovici {\em Nucl. Phys.} {\bf
  B509} (1998) 122, \href{http://xxx.lanl.gov/abs/hep-th/9707217}{{\tt
  hep-th/9707217}}.

\bibitem{Banks:1997vh}
T.~Banks, W.~Fischler, S.~H. Shenker, and L.~Susskind {\em Phys. Rev.} {\bf
  D55} (1997) 5112, \href{http://xxx.lanl.gov/abs/hep-th/9610043}{{\tt
  hep-th/9610043}};
L.~Susskind \href{http://xxx.lanl.gov/abs/hep-th/9704080}{{\tt
  hep-th/9704080}};
A.~Sen {\em Adv. Theor. Math. Phys.} {\bf 2} (1998) 51, 1998, 
\href{http://xxx.lanl.gov/abs/hep-th/9709220}{{\tt
    hep-th/9709220}}~;
N.~Seiberg {\em Phys. Rev. Lett.} {\bf 79} (1997) 3577
  \href{http://xxx.lanl.gov/abs/hep-th/9710009}{{\tt hep-th/9710009}}.

\bibitem{Blau:1997du}
M.~Blau and M.~O'Loughlin {\em Nucl. Phys.} {\bf B525} (1998) 182,
  \href{http://xxx.lanl.gov/abs/hep-th/9712047}{{\tt hep-th/9712047}};
C.~M. Hull {\em J. High Energy Phys.} {\bf 9807} (1998) 018,
  \href{http://xxx.lanl.gov/abs/hep-th/9712075}{{\tt hep-th/9712075}}.

\bibitem{Dijkgraaf:1997hk}
R.~Dijkgraaf, E.~Verlinde, and H.~Verlinde {\em Nucl. Phys.} {\bf B486} (1997)
  89, \href{http://xxx.lanl.gov/abs/hep-th/9604055}{{\tt hep-th/9604055}}.

\bibitem{Susskind:1996uh}
L.~Susskind \href{http://xxx.lanl.gov/abs/hep-th/9611164}{{\tt
  hep-th/9611164}};
O.~J. Ganor, S.~Ramgoolam, and W.~{Taylor IV} {\em Nucl. Phys.} {\bf B492}
  (1997) 191, \href{http://xxx.lanl.gov/abs/hep-th/9611202}{{\tt
  hep-th/9611202}}.

\end{thebibliography}
\end{document}